\documentclass[preprint,12pt]{elsarticle}

\usepackage{graphicx}
\usepackage{bm}
\usepackage{xcolor}
\usepackage{orcidlink}
\usepackage{booktabs}
\usepackage{amsmath,amssymb}
\usepackage[mathlines]{lineno}
\usepackage{hyperref}

\biboptions{sort&compress}

\journal{Computational Materials Science}

\begin{document}
\begin{frontmatter}

\title{Path-Resolved Molecular Dynamics Framework for Dumbbell-Interstitial Transport in Fe--Cr Alloys}

\author[nist,umd]{Yasir Mahmood\corref{cor}}
\ead{yasir.mahmood@nist.gov}

\author[inl]{Mathew M. Swisher}

\author[inl]{Sourabh Bhagwan Kadambi}

\author[inl]{Mukesh Bachhav}

\author[inl]{Jia-Hong Ke\corref{cor}}
\ead{jiahong.ke@inl.gov}

\cortext[cor]{Corresponding authors.}

\affiliation[nist]{
organization={Materials Measurement Laboratory, National Institute of Standards and Technology},
city={Gaithersburg},
state={MD},
postcode={20899},
country={USA}
}

\affiliation[umd]{
organization={Department of Chemistry \& Biochemistry, University of Maryland},
city={College Park},
state={MD},
postcode={20742},
country={USA}
}

\affiliation[inl]{
organization={Idaho National Laboratory},
city={Idaho Falls},
state={ID},
postcode={83415},
country={USA}
}

\begin{abstract}
Accurate prediction of radiation-induced segregation in structural alloys hinges on the availability of species-specific transport coefficients, particularly the defect-solute coupling terms that govern correlated atomic transport. While molecular dynamics (MD) simulations converge relatively quickly for species self-transport coefficients, cross-species coupling terms remain challenging to evaluate owing to slow statistical convergence. This challenge becomes even more significant in non-dilute or concentrated alloys, where transport is influenced by a diverse range of local chemical environments that are difficult to represent within dilute-alloy kinetic descriptions. In this work, we present a path-resolved MD framework for calculating self-interstitial-mediated transport in model Fe--Cr alloys. Our approach focuses on atoms lying along the defect trajectory and dynamically disturbed by defect motion, rather than averaging over the entire system, thereby enabling improved statistical reliability in the calculation of both diagonal and off-diagonal transport coefficients. We further track the evolving chemical identity of the migrating dumbbell and show a pronounced increase in Fe--Fe dumbbell occupancy at elevated temperatures in dilute alloys, accompanied by a weakening of Cr defect displacement coupling at elevated temperatures. These configuration and coupling trends provide a consistent microscopic basis for the temperature-composition dependence of the defect-mediated Onsager coefficients and the resulting Cr selectivity of interstitial transport. This work demonstrates the potential of path-resolved MD as a practical route for evaluating interstitial-mediated transport in concentrated alloys and informing multiscale models of radiation-induced segregation.
\end{abstract}


\end{frontmatter}


\section{\label{sec:intro}Introduction}
In binary alloy systems, the long-range solute transport is predominantly mediated via point defects. The defect-driven diffusion proceeds either via atom-vacancy exchange or by self-interstitial hopping. These mechanisms become more dominant under irradiation, where vacancies and self-interstitial atoms are produced continuously and in far excess of their thermal equilibrium concentrations, contributing significantly more to transport driven microstructural evolution such as radiation-induced segregation (RIS)~\cite{nastar20121,wiedersich1979theory} of solute at sinks like grain boundaries (GBs)~\cite{okamoto1974segregation, wharry2013systematic,kadambi2025phase} and dislocation loops~\cite{barbu1975irradiation,KE201883}. When the solute segregation reaches the solubility limit, it can further cause radiation-induced precipitation~\cite{barbu1975irradiation, ke2022cluster,KE201883, meslin2013radiation,ke2023microstructure} which can raise technological concerns such as corrosion, embrittlement, and cracking~\cite{jiao2011novel, nakata1987solute}.

Under thermal equilibrium, diffusion is typically mediated by vacancies because equilibrium self-interstitial concentrations are negligible compared to vacancies~\cite{flynn1972point, dienes1965point}, despite higher interstitial mobility~\cite{olsson2007ab}. However, irradiation can tip this scale when interstitials and vacancies are generated in comparable numbers, making interstitial-mediated transport significant and allowing it to compete directly with vacancy-mediated transport in driving coupled solute fluxes, depending on alloy composition and temperature. This competing effect is critical to RIS, as evidenced by several multiscale atomistic studies which show that vacancy and interstitial mechanisms can bias solute transport in opposite directions, leading to either solute enrichment or depletion at sinks depending on external stimuli and alloy chemistry~\cite{choudhury2011ab, senninger2016modeling, piochaud2016atomic}. Choudhury et al. combined first-principles energetics with statistical-mechanics diffusion models to compute the full Onsager transport matrix and tracer diffusivities for both vacancy-mediated and dumbbell-mediated transport in dilute BCC Fe--Cr. They showed that the off-diagonal or coupled coefficients are often comparable in magnitude to the diagonal terms for both mechanisms, vacancy-mediated and interstitial-mediated transport. Based on the competing effects between the two under a given set of conditions, the RIS tendencies for Cr (depletion vs enrichment) can be predicted~\cite{choudhury2011ab}. Building on this picture, Senninger et al. employed an atomistic kinetic Monte Carlo framework parameterized by density functional theory (DFT) that explicitly includes vacancy and self-interstitial diffusion, recombination, and elimination at sinks, and used it to quantify how the relative vacancy and interstitial contributions evolve with temperature and composition. In particular, they found that the relative preference of each mechanism decreases with increasing temperature, with the interstitial contribution dominating below a composition-dependent threshold temperature and the vacancy contribution dominating above it. Their results provide a mechanistic basis for why RIS trends can invert as irradiation conditions and alloy chemistry change \cite{senninger2016modeling}.

While these studies emphasize the competition between vacancy and interstitial mediated fluxes, the interstitial mechanism is itself configuration dependent. For interstitial-mediated transport in a binary alloy with species A and B, the migrating dumbbell can switch between A-A, A-B, and B-B states, and the residence of these states and the associated trapping kinetics control how strongly the solute displacement is correlated with the interstitial motion, and therefore how influential the solute-defect coupling terms in the transport matrix can be. While prior work has established that the solute-interstitial interactions and the stability of particular dumbbell configurations are strongly concentration dependent, direct evaluation of the transport couplings generated by these interactions remains challenging, particularly in concentrated alloys where migrating interstitials encounter a broad range of local chemical environments. This motivates the use of molecular dynamics (MD) as a complementary approach because the dumbbell migration barriers are low enough to capture defect motion on MD accessible timescales, enabling direct sampling of configuration switching, residence times, and the correlated displacement response that govern the off-diagonal Onsager transport coefficients. In this work, we use MD to directly connect dumbbell configuration evolution with interstitial-mediated transport coupling in Fe--Cr alloys while introducing a path-resolved framework for evaluating off-diagonal transport coefficients.

In principle, $L_{ij}$ can be extracted directly from atomistic trajectories using the generalized Einstein formulation for the Onsager matrix~\cite{einstein1905motion, allnatt1982einstein}, but in practice, the off-diagonal terms are notoriously slow to converge~\cite{piochaud2016atomic}. Barnard and Morgan, using ab initio methods, explicitly noted that off-diagonal coefficients are particularly slow to converge and require substantially longer trajectories than are typically feasible in atomistic simulation time scales~\cite{barnard2014ab}. Moreover, when using a standard MD protocol for these calculations, such as microcanonical ensemble (NVE) with periodic boundary conditions, the absence of external forces implies conservation of the total linear momentum ($P$) of the system,
\begin{equation}
\label{eq:mom_conserved}
\frac{d\mathbf{P}}{dt}=\frac{d}{dt}\sum_{k=1}^{N} m_k \mathbf{v}_k = \mathbf{0},
\end{equation}
where $N$ is the total number of atoms and $m_k$ and $\mathbf{v}_k$ are the mass and velocity of atom $k$. If the initial net momentum is set to zero, as is typically enforced during initialization, the center of mass (COM) velocity remains zero and the COM position is stationary up to numerical round off. In a finite periodic crystal, when the defect translates, it causes a long ranged elastic response of the surrounding lattice that is observed as a counteracting translational collective displacement field. 

For the diagonal terms $L_{ii}$ of the Onsager matrix, this background contribution is generally less problematic because the self terms are dominated by the species mean-square displacement. By contrast, the off-diagonal coefficients $L_{ij}$ with $i \neq j$ depend on the covariance between two species displacements, so even a small coherent background displacement can contribute to $\langle \Delta \mathbf{R}_{i}(t)\cdot \Delta \mathbf{R}_{j}(t) \rangle$, where $i$ and $j$ denote atomic species. This can dominate the covariance even if the background drift is weak and can contaminate the cross correlations used to evaluate $L_{ij}$ (for $i\ne j$). This sensitivity motivates a path-resolved MD strategy that isolates the localized displacement field associated with defect motion before evaluating the off-diagonal transport couplings.
\begin{figure}[t!]
    \begin{center}
        \begin{tabular}{c}
            \includegraphics[width = 0.99\textwidth]{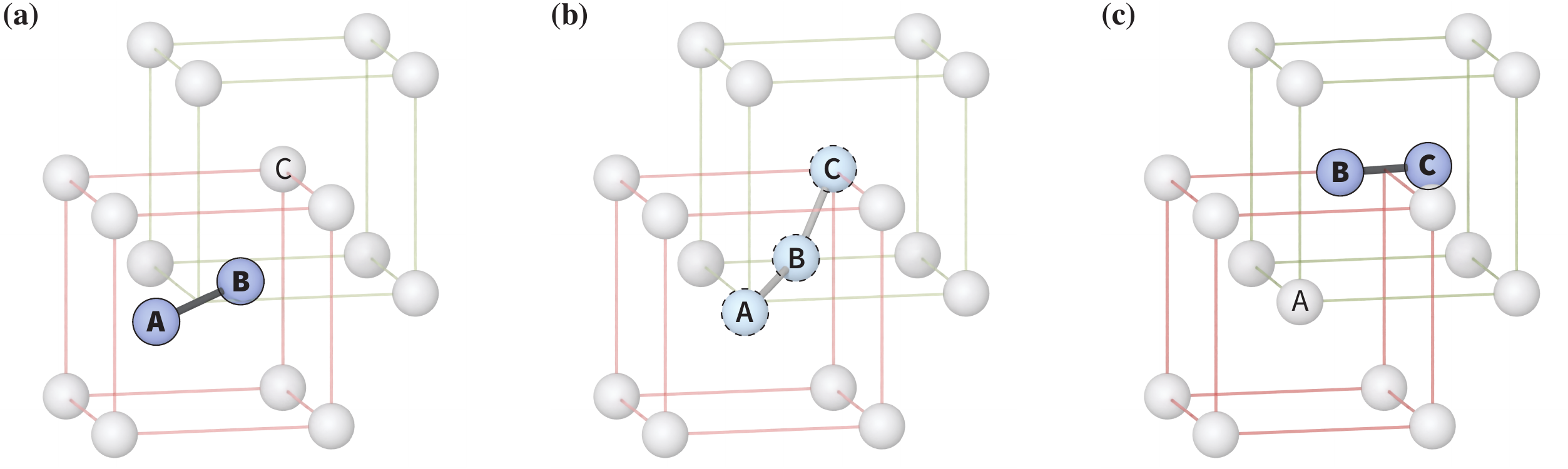}
        \end{tabular}
    \end{center}
    \caption{\label{fig:dI_schematic} Dumbbell interstitial migration between neighboring lattice sites (schematic), with bonds shown for perspective purposes. All lattice site atoms are shown as gray spheres. The two atoms forming the dumbbell are highlighted in blue while atoms participating in the exchange during the transition are shown in light blue. (a) Initial dumbbell composed of atoms A and B. (b) Transition configuration in which the leading atom B engages neighboring atom C, producing a transient three-atom configuration (A–B–C) during hopping to the adjacent site. (c) Final configuration after transition, where the dumbbell has moved and is now formed by atoms B and C.}
\end{figure}
\section{\label{sec:methods}Methods}
\subsection{Simulation setup}
Dumbbell-mediated self-interstitial migration proceeds through a sequence of interstitialcy or bond-switching events, distinct from the atom-vacancy exchange process that controls vacancy diffusion. During a hop, the leading atom of the dumbbell pair forms a transient interaction with a neighboring lattice atom while the original dumbbell bond progressively weakens~\cite{fu2004stability, messina2020solute}. This intermediate configuration is schematically illustrated in Fig.~\ref{fig:dI_schematic}, where two adjacent BCC unit cells are shown to highlight the local exchange geometry. Initially, the dumbbell interstitial is formed by atoms A and B within the first unit cell, shown in light red. As atom B begins to interact with atom C, the A-B bond is destabilized and the interstitial defect advances through a bond-switching event, ultimately forming a new B-C dumbbell at the neighboring lattice site. Depending on the local pathway, this migration event may involve both translational and rotational components~\cite{barbe2007phenomenological, choudhury2011ab, messina2020solute, vincent2006ab}. This exchange sequence provides the microscopic basis for defining the atoms that participate directly in dumbbell transport. A representative dumbbell hop from the MD trajectory is provided in the Supplementary Material as a video.

The MD simulations were performed using the Large-scale Atomic/Molecular Massively Parallel Simulator (LAMMPS)~\cite{LAMMPS} in three dimensions with periodic boundary conditions. Certain software packages are identified in this paper solely to describe the procedures used. Such identification does not imply recommendation or endorsement by the National Institute of Standards and Technology (NIST), nor does it imply that the software identified is necessarily the best available for the purpose. A BCC Fe supercell was constructed with an initial lattice parameter $a_0=2.86$~\AA~and a linear size of $20~a_0$ along each $\langle100\rangle$ direction. The interatomic interactions were modeled using a Fe--Cr Embedded Atom Method (EAM)~\cite{daw1984embedded} based alloy potential~\cite{bonny2011iron} developed for high-chromium ferritic alloys. Since defect energetics can be sensitive to the choice of interatomic potential~\cite{mahmood2024universal}, the transport coefficients reported here should be interpreted within the limits of this potential. A single self-interstitial dumbbell was introduced at the cell center by removing the lattice atom at the insertion site and placing two Fe atoms displaced by $\pm 0.05\,a_0$ about the center (initially along a random direction). The resulting configuration was then relaxed at $0$~K using conjugate-gradient energy minimization~\cite{polak1969note}, during which the dumbbell adopts the stable $\langle110\rangle$ orientation~\cite{fu2004stability, vincent2006ab}.

A small random perturbation was applied to atoms within a spherical region around the defect core to avoid symmetry locking. Cr was then introduced by random substitution of Fe sites to generate alloys with Cr compositions of 1, 3, 6, and 9 at.\%, followed by further structural relaxation. Each composition was equilibrated at the target temperatures of 500~K, 600~K, 700~K, 800~K, and 900~K using an isothermal-isobaric (NPT) ensemble at zero pressure. Subsequently, production trajectories were generated in the microcanonical (NVE) ensemble. For each composition and temperature state point, 10 statistically independent runs were performed using different random seeds. Each production simulation was 300~ns long, and defect and atomic displacements were recorded every 1~ns~(or every 1~million timesteps). Over a typical 1~ns sampling interval, the dumbbell executed approximately 1500 to 2000 hops, providing sufficient statistics for computing ensemble-averaged transport coefficients.

\subsection{Trajectory/Non-trajectory method}
During the production stage, defect migration was simulated in the NVE ensemble under periodic boundary conditions. In this setting, the forward motion of the dumbbell is accompanied by compensating displacements of the surrounding lattice due to total momentum conservation and long-range elastic relaxation of the crystal, as described by Eq.~(\ref{eq:mom_conserved}). These effects introduce a long wavelength and system-wide opposing displacement field. As a consequence, the collective displacements of atoms used to evaluate cross-correlation transport coefficients can contain a significant background contribution that is not directly associated with the local defect-transport mechanism, particularly in finite simulation supercells. This issue is particularly critical when quantifying coupled solute-defect transport, as such background contribution can lead to significant errors in the off-diagonal components of the Onsager transport coefficient matrix, which are known to be very sensitive to correlated atomic displacements in MD simulations.

\begin{figure}[t!]
    \begin{center}
        \begin{tabular}{c}
            \includegraphics[width = 0.99\textwidth]{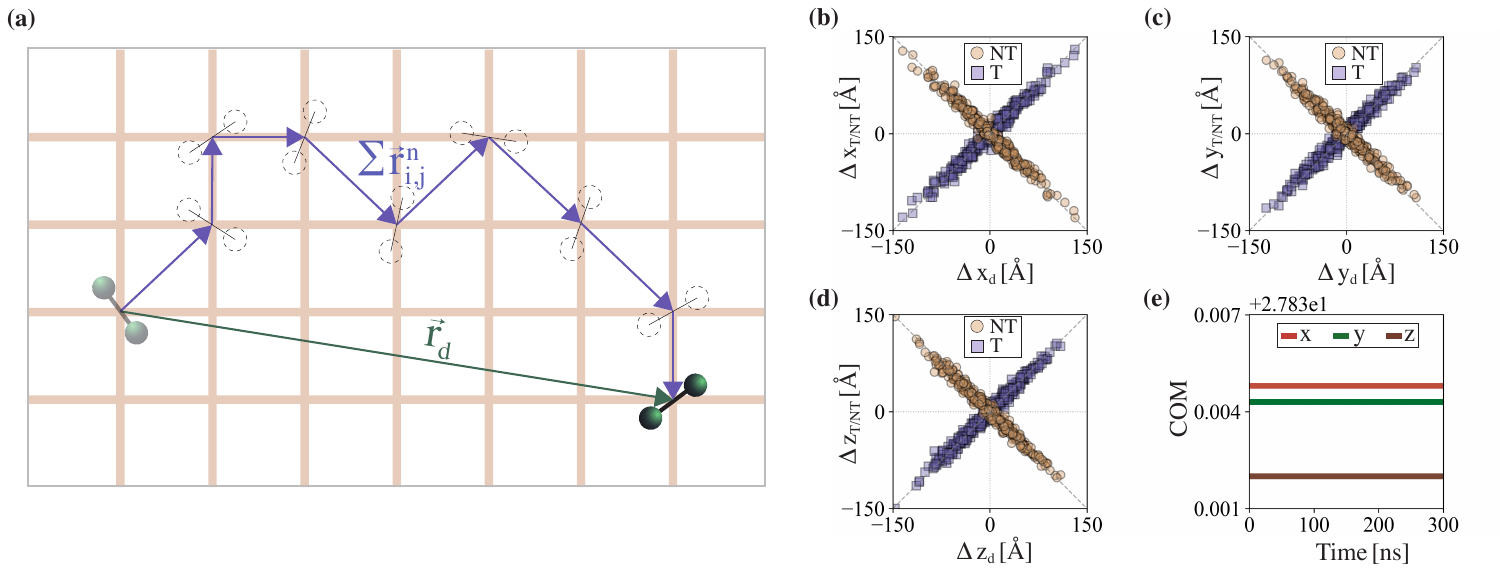}
        \end{tabular}
    \end{center}
    \caption{\label{fig:fig_TNT} (a) Schematic of the dumbbell translation from lattice site 1 to 2 via successive atom-exchange events along the migration pathway. Atoms located at lattice sites along the defect trajectory (shown in purple) that participate in the exchange sequence are defined as trajectory atoms. All other atoms are classified as non-trajectory atoms (not shown for clarity). (b)-(d) Component-wise displacement of the trajectory and non-trajectory atom sets, $\Delta x$, $\Delta y$, and $\Delta z$, plotted against the corresponding dumbbell displacement components, $\Delta x_d$, $\Delta y_d$, and $\Delta z_d$. (e) Center of mass displacement of all the atoms, including the dumbbell.}
\end{figure}

To mitigate this background contribution, we restrict the displacement analysis to atoms that participate directly in the defect migration trajectory, as illustrated in Fig.~\ref{fig:fig_TNT}(a). The dumbbell configuration at the reference time $t=0$ is shown in transparent green, while the configuration at time $t$ is shown in green. Successive migration events define a discrete path through the lattice (purple arrows). Atoms belonging to the sequence of sites visited and exchanged during dumbbell translation are grouped as trajectory (T) atoms, for which the cumulative displacement is directly related to the net defect translation, $\sum_{n}\vec{r}^{~n}_{i} = \vec{r}_{d}$, where $i$ spans the Fe and Cr atoms involved along the defect path. All remaining atoms that are not associated with the path are grouped as non-trajectory (NT) atoms and are treated as the background lattice response. The NT atoms are excluded from the species collective displacement sums and are used to define the material reference frame. In BCC Fe, a single dumbbell hop advances the leading atom by approximately one half of the nearest-neighbor distance, $\approx 1.24$~\AA. Accordingly, we identify trajectory atoms operationally as those that (i) lie along the defect migration path and (ii) undergo a displacement magnitude exceeding $1$~\AA\ within a sampling interval.

Figs.~\ref{fig:fig_TNT}(b)-(d) validate the T/NT decomposition at the level of the recorded displacement signals from one of the simulations. Each panel shows the dumbbell interstitial (dI) displacement component, $\Delta x_{dI}$, $\Delta y_{dI}$, or $\Delta z_{dI}$, plotted against the corresponding collective displacement of the T atoms (purple squares) and the NT atoms (orange circles), with one point recorded per sampling interval (every 1~ns). In all three components, the T-atoms' displacement signal tracks the dumbbell translation with a near-linear, near-unity correlation, consistent with T atoms carrying the localized displacement associated with the bond-exchange sequence. By contrast, the NT atoms' displacement exhibits a pronounced anti-correlation, reflecting the compensating displacement field of the surrounding lattice. Fig.~\ref{fig:fig_TNT}(e) shows that the system COM position remains essentially constant (within numerical noise), indicating that the anti-correlated NT response corresponds to a compensating background field rather than a net translation of the simulation cell.
\subsection{Transport coefficient calculations}
For the transport calculations we employed the generalized Einstein formulation for the Onsager matrix~\cite{allnatt2003atomic}:
\begin{equation}
\label{eq:gen_eins}
L_{ij} = \frac{\langle\Delta \mathbf{R}_i \cdot \Delta \mathbf{R}_j\rangle}{6V_{sys}k_{B}T\Delta t}
\end{equation}
where $\Delta \mathbf{R}_i$ is the collective displacement vector of T atoms with respect to NT atoms (material frame of reference) of that species $i$ in time $\Delta t$, $V_{sys}$ is the system volume, $k_B$ is the Boltzmann constant, $T$ is the temperature. For simplicity, we define the normalized Onsager matrix for dumbbell interstitial-mediated transport as~\cite{piochaud2016atomic},
\begin{equation}
\label{eq:gen_eins_norm}
l^{dI}_{ij} = \frac{V_{at}k_{B}T}{X_{dI}}L_{ij}
\end{equation}
where $V_{at}$ is the atomic volume (lattice-site volume), and $X_{dI}$ is the dumbbell interstitial concentration, which for a single dumbbell in the simulation cell is $X_{dI}=1/N_{\mathrm{sites}}$. In evaluating the collective displacements entering Eq.~(\ref{eq:gen_eins}), and therefore the normalized coefficients in Eq.~(\ref{eq:gen_eins_norm}), the displacements of the T atoms and the dumbbell are computed with respect to the NT atoms, which define the material reference frame.

\section{\label{sec:rnd}Results and Discussion}
\subsection{Dumbbell tracer diffusion coefficient}
The dumbbell interstitial diffusivity, $D_{dumbbell}$, was calculated using the Einstein relation for random walk~\cite{einstein1905motion, helfand1960transport}, 
\begin{equation}
\label{eq:D_eins}
D=\frac{\langle \Delta r^2 \rangle}{6t}
\end{equation}
using the net dumbbell displacements with respect to NT atoms and then averaged over 10 statistically independent trajectories at each (at.\% Cr and T) state point. Over the temperature range 500~K to 900~K, $D_{dumbbell}$ increases by nearly 2 orders of magnitude, from $(8.28-13.39)\times10^{-10}~m^2/s$ at 500~K to $(3.83-5.02)\times10^{-8}~m^2/s$ at 900~K, with only a modest variation across $1-9$ at.\% Cr. In an Arrhenius representation, the data fall on linear trends, consistent with thermally activated migration.
The linear fits yield activation energies, $E_a$, in the range $0.33-0.38$~eV (with $R^2>0.996$ for all compositions). The fitted activation energies and prefactors are summarized in Table~\ref{tab:arrhenius}. These magnitudes and activation energies are consistent, within the range of comparable states, with prior atomistic studies of self-interstitial mediated transport in Fe--Cr alloys, including kinetic Monte Carlo based treatments~\cite{senninger2016modeling}, MD simulations~\cite{terentyev2008migration}, and broadly are comparable to experimental measurements~\cite{kuvcera1974self}. Senninger et al.~\cite{senninger2016modeling} reported effective migration energies of approximately 0.29 and 0.27~eV at 5 and 10 at.\% Cr, respectively, while Terentyev et al.~\cite{terentyev2008migration} reported corresponding values of 0.28 and 0.25~eV.
\begin{figure}[t!]
    \begin{center}
        \begin{tabular}{c}
            \includegraphics[width = 0.5\textwidth]{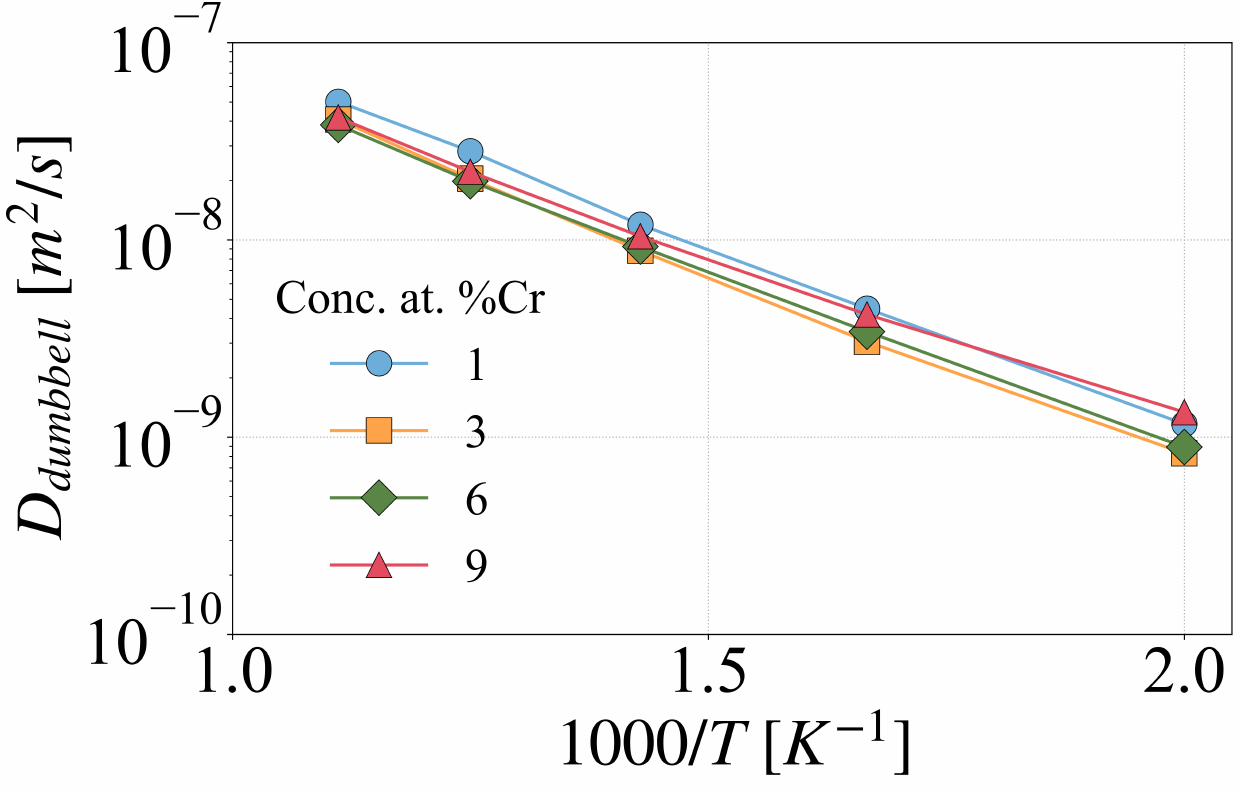}
        \end{tabular}
    \end{center}
    \caption{\label{fig:D_dIvsTemp} Diffusion coefficient for the dumbbell interstitial defect in Fe--Cr alloys at different temperatures and concentrations with respect to non-trajectory atoms (material frame of reference).}
\end{figure}

Note that $D_{dumbbell}$ shows only a modest dependence on Cr composition over the range $1-9$ at.\%, with the activation energy varying only slightly across compositions (Fig.~\ref{fig:D_dIvsTemp}), which is consistent with experimental measurements~\cite{kuvcera1974self}. This observation suggests that over the composition and temperature ranges investigated here, total interstitial diffusion is not strongly affected by Cr content. This behavior is consistent with previous mesoscale modeling of ferritic-martensitic Fe--Cr alloys, which often assumes a composition-independent activation energy for interstitial diffusivity~\cite{field2013dependence,KE201883,KE2019586}. Importantly, the weak composition dependence of the overall dumbbell diffusivity does not imply a similarly weak dependence of defect-mediated transport coupling. As shown in the following sections, strong variations in dumbbell chemistry, species-defect correlations, and off-diagonal transport coefficients can occur even when the overall interstitial mobility remains relatively unchanged.
\begin{table}[b]
\centering
\caption{Arrhenius parameters for dumbbell-interstitial diffusion.}
\label{tab:arrhenius}
\begin{tabular}{ccc}
\toprule
Cr content (at.\%) & $E_a$ (eV) & $D_0$ ($10^{-6}$ m$^2$/s) \\
\midrule
1 & 0.366 & 5.49 \\
3 & 0.378 & 4.91 \\
6 & 0.363 & 3.89 \\
9 & 0.332 & 2.77 \\
\bottomrule
\end{tabular}
\end{table}
\subsection{Dumbbell configurations and correlations}
\begin{figure}[t!]
    \begin{center}
        \begin{tabular}{c}
            \includegraphics[width = 0.99\textwidth]{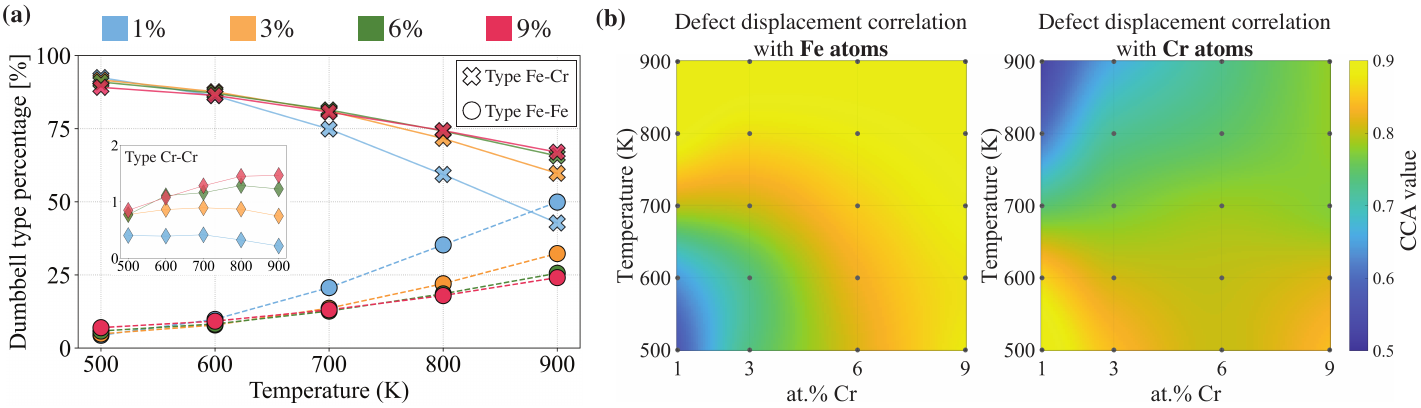}
        \end{tabular}
    \end{center}
    \caption{\label{fig:dIevolution} (a) Residence fractions of the dumbbell interstitial in Fe--Fe, Fe--Cr, and Cr--Cr configurations as functions of temperature and at.\% Cr. (b) Canonical correlation analysis (CCA) quantifying the 3D displacement coupling between dumbbell translation and the trajectory-atoms displacement, evaluated separately for Fe and Cr. Gray markers indicate the mean over independent runs at each state point $(c_{Cr},~T)$ and the surface is a linear interpolation of the grid.}
\end{figure}

The dumbbell interstitial can be classified into three different types, Fe--Fe, Fe--Cr, and Cr--Cr based on the chemical identity of the two atoms forming the defect. Across the different states investigated, the defect resides predominantly in the mixed Fe--Cr configuration at low temperatures. At 500~K, the mixed type accounts for 85~\% to 90~\% of the total trajectory, while the Fe--Fe and Cr--Cr configurations contribute $<15~\%$ and $<1~\%$ of the trajectory, respectively, Fig.~\ref{fig:dIevolution}(a). As the temperature increases, the preferred configuration shifts from Fe--Cr towards Fe--Fe. The effect is more pronounced in dilute alloys, where at 1 at.\% Cr, the Fe--Fe percentage increases to $\approx50~\%$ by 900 K (with Fe--Cr correspondingly reduced to $\approx45~\%$), whereas at higher Cr concentrations ($3-9$ at.~\%) Fe--Cr remains the majority species even at 900~K, $\approx65~\%$ of the total simulation time, and Fe--Fe remains at $\approx35~\%$. One important point overlooked by the configuration percentages is the residence time asymmetry between configurations. The Fe--Cr dumbbell, although present for a significant fraction of the simulation time, exhibits substantially longer dwell times than Fe--Fe, so once the migrating defect encounters a Cr atom it can remain in a low-energy mixed dumbbell environment for extended periods before continuing its migration. This interpretation is consistent with the statistical-mechanics argument emphasized by Terentyev et al. that the time spent in a low-energy configuration exceeds that in a higher-energy configuration by a factor proportional to $\exp(\Delta E/k_{\mathrm{B}}T)$. They further show that configurations involving more than one Cr atom can trap the dumbbell~\cite{terentyev2008migration}.

\par %
To better highlight the dumbbell's role in defect migration, we assess the extent to which Fe and Cr displacements are dynamically coupled to defect translation. We perform a canonical correlation analysis (CCA)~\cite{hotelling1992relations, hardoon2004canonical}, which combines the three Cartesian displacement components into a single measure of dumbbell-species displacement coupling, Fig.~\ref{fig:dIevolution}(b). The gray markers show the data at each (at.\% Cr and temperature) state point, plotted as the mean over the independent runs. The surface is a visualization aid obtained by linear interpolation of the gridded means. For Fe, the dumbbell-Fe coupling is high over most of the concentration and temperature space (typically $\approx$~0.8 to 0.9), with the only notable reduction occurring in the dilute/low-temperature corner where the mean CCA value drops to $\approx0.6$. In contrast, Cr shows the opposite trend, the correlation is strongest at low concentrations and low temperatures, $\approx~0.92$ at 1 at.\% Cr and 500~K, but weakens systematically with increasing temperature in the dilute limit, reaching its minimum at 1 at.\% Cr and 900~K ($\approx~0.52$), while remaining moderate at higher Cr concentrations, for example, $\approx~0.66-0.78$ at 900~K for $3-9$ at.\% Cr. These trends are consistent with dumbbell type statistics. At low temperature, the defect spends most of its time as a mixed Fe--Cr dumbbell, so Cr motion remains strongly coupled to defect translation. As temperature increases, the fraction of Fe--Fe dumbbells grows, particularly in the dilute limit, reducing the relative participation of Cr in defect migration and weakening the Cr-dumbbell coupling. At higher Cr concentrations and elevated temperatures, Fe--Cr remains the dominant dumbbell configuration, although the residence fraction of Fe--Fe dumbbells increases appreciably. Consequently, the decrease in the Cr–defect displacement correlation is more moderate than in dilute alloys.

The CCA should not be interpreted as biased to Fe owing to the higher number of atoms. In this implementation, the CCA is performed on the three signed displacement components after centering and whitening, and is therefore insensitive to an overall scale difference in displacement magnitude. Accordingly, a larger cumulative Fe displacement does not by itself produce a larger correlation with the dumbbell trajectory. Rather, the CCA quantifies the degree to which the dumbbell displacement and the species-resolved displacement field evolve in time and direction. Any observed Fe--Cr difference therefore reflects differences in the structure and coherence of the displacement response associated with defect migration.

\subsection{Onsager coefficients}
\begin{figure}[t!]
    \begin{center}
        \begin{tabular}{c}
            \includegraphics[width = 0.99\textwidth]{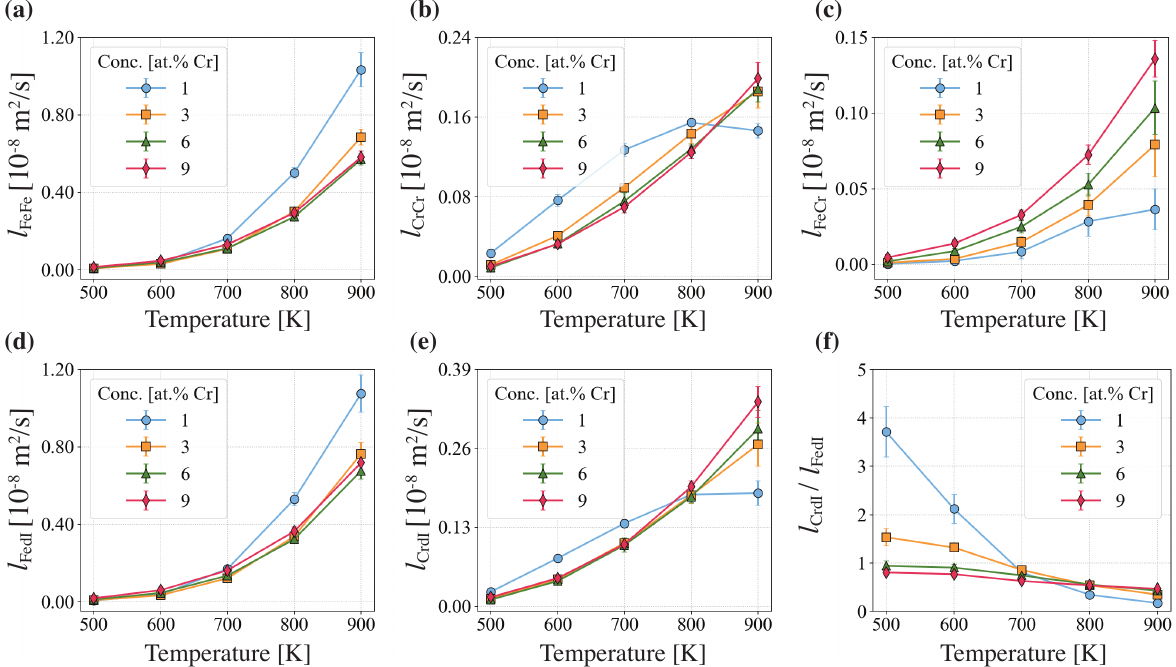}
        \end{tabular}
    \end{center}
    \caption{\label{fig:Lij} Dumbbell interstitial-mediated normalized Onsager transport coefficients in Fe--Cr: (a) $l_{FeFe}$, (b) $l_{CrCr}$, (c) $l_{FeCr}$, (d) $l_{FedI}$, (e) $l_{CrdI}$, and (f) $l_{CrdI}/l_{FedI}$, shown as functions of temperature and at.\% Cr concentration. Coefficients are computed from Eq.~(\ref{eq:gen_eins_norm}) using trajectory-atom displacements referenced to the non-trajectory atom frame (or material frame of reference).}
\end{figure}

For the Onsager coefficient calculations using Eq.~(\ref{eq:gen_eins_norm}), the collective displacement vectors of the dumbbell and the trajectory atoms were computed with respect to the frame of reference of non-trajectory atoms, thereby removing the background lattice response associated with momentum conservation and long-ranged elastic relaxation. Fig.~\ref{fig:Lij} shows the defect-mediated Onsager coefficients as functions of temperature and concentration, with markers representing averages over ten independent simulations and error bars denoting the standard deviation across runs. The resulting temperature and composition dependent trends are consistent with the dumbbell configuration statistics and CCA coupling maps in Fig.~\ref{fig:dIevolution}. For comparison, Onsager coefficients computed using all-atom displacements in the laboratory frame, without the T/NT decomposition, are provided in the Supplementary Material.

$l_{FeFe}$ increases monotonically with temperature in all states with the strongest temperature dependence observed in the dilute alloy (1 at.\% Cr). This trend is reflective of the pronounced shift in defect configuration occupancy from Fe--Cr towards Fe--Fe at high temperatures in the dilute limit, which consequently increases the migration of Fe within the lattice. This observation is further bolstered by the CCA maps in Fig.~\ref{fig:dIevolution}(b), which show that Fe remains strongly coupled to dumbbell translation at higher temperatures.
The $l_{CrCr}$ response is more nuanced because Cr participation in the migration pathway depends primarily on the availability of mixed Fe--Cr dumbbells (since Cr--Cr configuration remains minuscule in every state, Fig.~\ref{fig:dIevolution}(a) inset) and the degree of Cr-dumbbell dynamical coupling. In the dilute case of 1 at.\% Cr, the $l_{CrCr}$ exhibits a maximum value near 800~K followed by reduction at 900~K, consistent with the onset of weaker Cr-dumbbell coupling at high temperatures. The increasing Fe--Fe configuration occupancy at higher temperatures further curbs the Cr atom involvement in bond-exchange and hopping sequence, therefore suppressing the Cr transport. As the Cr content increases to 3 at.\%, this turnover is diminished, and at 6 and 9 at.\%, the coefficient increases monotonically with temperature, in line with persistence of Fe--Cr as the dominant dumbbell type and the moderate defect displacement correlation at higher T. Finally, the cross coefficient $l_{FeCr}$ shows the strongest temperature growth at higher Cr concentrations, reflecting both the larger fraction of time spent in mixed Fe--Cr configurations and the sustained dynamical coupling between the Cr motion and dumbbell translation in that regime. In contrast, at 1 at.\% Cr, the increase in $l_{FeCr}$ begins to plateau above 800~K as Fe--Fe migration becomes more prevalent and Cr becomes progressively less locked to defect motion, as indicated by the CCA map.

The defect-species cross coefficients further classify how Fe and Cr participate in dumbbell mediated transport. The coefficient $l_{FedI}$ increases monotonically with temperature for all compositions, with the steepest rise observed at 1 at.\% Cr. This trend is consistent with the increasing participation of Fe in the migration pathway at elevated temperature, particularly in the dilute limit where the defect residence shifts strongly from Fe--Cr towards Fe--Fe configurations. In parallel, the CCA results showed that Fe remains strongly coupled to dumbbell translation across the full temperature and composition range investigated. Hence, the increase in $l_{FedI}$ reflects both thermal activation of defect motion and enhanced Fe involvement in the bond switching mechanisms.

The $l_{CrdI}$ exhibits a stronger composition dependence. Although $l_{CrdI}$ also increases with temperature at most compositions, the dilute limit shows a tendency towards saturation. At 1 at.\% Cr, the rise plateaus after 800~K, consistent with the reduced Cr-dumbbell dynamical coupling seen in the CCA map and with the decreasing residence of the mixed type dumbbell at higher temperatures. At higher Cr contents, however, this plateauing behavior is suppressed and $l_{CrdI}$ rises more steadily with temperature, with the strongest increase observed at 9 at.\% Cr. This reflects the greater availability of mixed type dumbbell configurations and the moderate participation in the defect migration pathway even at elevated temperatures.

The ratio $l_{CrdI}/l_{FedI}$ in Fig.~\ref{fig:Lij}(f) provides a measure of the interstitial-driven selectivity of Cr relative to Fe and shows how strongly Cr displacements remain locked to dumbbell translation compared with Fe. The ratio decreases with temperature for all compositions and the degree of decrease gets weaker as the bulk Cr content increases. In the dilute alloy (1 at.\% Cr), $l_{CrdI}/l_{FedI}$ drops sharply from $\approx 3.7$ at 500~K to $\approx 0.17$ at 900~K. The observation is consistent with the stronger predominance of mixed Fe--Cr dumbbells and the more moderate reduction in Cr-dumbbell coupling at higher Cr concentrations.

\subsection{Radiation-induced segregation implications}
The Onsager coefficient trends observed qualitatively align with RIS studies that attribute Cr enrichment at sinks at lower temperature to stronger coupling of Cr with self-interstitial transport and report that the magnitude of Cr enrichment decreases with increasing bulk Cr content in ferritic/martensitic alloys, with enrichment weakening as the temperature climbs. In particular, Pechenkin et al. combined MD-based characterization of interstitial-mediated transport in Fe--Cr and predicted that the degree of enrichment increases as the bulk Cr content and temperature decrease when the interstitial mechanism is dominant \cite{pechenkin2013radiation}. Experimentally, Wharry and Was performed a systematic irradiation study on commercial ferritic-martensitic steels and quantified RIS at prior-austenite GBs using scanning transmission electron microscopy/energy dispersive X-ray spectroscopy (STEM/EDX). They showed that under fixed irradiation conditions, the extent of Cr enrichment (accompanied with Fe depletion) decreases with increasing bulk Cr concentration and temperature~\cite{wharry2013systematic}. Senninger et al. reached similar conclusions using a DFT parameterized atomistic kinetic Monte Carlo framework that included both vacancy and self-interstitial diffusion. They showed that the net Cr enrichment/depletion tendency is governed by the balance of the two defect contributions. For the interstitial, they showed the ratio $L_{Cr,I}/L_{Fe,I}$ weakens with temperature implying that the relative Cr-dumbbell coupling weakens more strongly than the corresponding Fe-dumbbell coupling at elevated temperatures~\cite{senninger2016modeling}. 

While these studies primarily interpret RIS in terms of the competition between vacancy and interstitial mediated fluxes, our configuration resolved analysis provides a complementary perspective on the interstitial channel itself. Specifically, resolving the dumbbell chemistry shows that the increased occupancy of Fe--Fe dumbbells at elevated temperatures, particularly in dilute alloys, shift the correlated displacement associated with interstitial migration to become increasingly Fe dominated and reduces Cr-selective transport. At higher Cr concentrations, mixed Fe--Cr dumbbells remain sufficiently prevalent at elevated temperature to sustain Cr participation in interstitial-mediated transport. Therefore, the reduction in Cr–dumbbell coupling is more gradual than in dilute alloys, where the increasing Fe–Fe fraction drives a stronger shift toward Fe-dominated transport. The strongest interstitial contribution to Cr enrichment is therefore expected in the low temperature dilute regime, where Cr remains more strongly coupled to defect migration relative to Fe, as reflected by the larger $l_{CrdI}/l_{FedI}$ ratios (Fig.~\ref{fig:Lij}(f)) and stronger Cr-defect displacement correlations (Fig.~\ref{fig:dIevolution}(b)) observed at low Cr concentrations.

More broadly, these results suggest a route for improving mesoscale RIS or phase-field models that treat interstitial transport using composition and temperature dependent effective Onsager coefficients~\cite{Oplinger2026}. Rather than representing interstitial mediated transport with a single averaged coefficient, configuration-resolved inputs could be used to weight the contributions from Fe--Fe, Fe--Cr, and Cr--Cr dumbbells according to their residence fractions and species-specific displacement correlations. Such a parameterization would allow phase-field models to retain information about the dominant dumbbell configuration and its associated transport selectivity, thereby connecting atomistic defect chemistry more directly to continuum scale segregation predictions. The present results therefore provide a microscopic explanation for why the interstitial contribution to Cr enrichment is expected to be strongest in the dilute and low temperature regime and to diminish progressively as temperature increases.

\section{\label{sec:conclusion}Conclusions}
In conclusion, we introduced a path-resolved MD framework for recording defect and atomic displacements that enables evaluation of both diagonal and off-diagonal transport coefficients for interstitial-mediated diffusion in a model Fe--Cr alloy. By computing collective displacements of the dumbbell and trajectory atoms in a reference frame defined by the non-trajectory atoms, the approach suppresses the background response associated with momentum conservation and elastic relaxation in finite periodic system, improving the statistical reliability of cross-correlation transport terms. Using this framework, we quantified dumbbell tracer diffusion over 500~K to 900~K and found an approximately Arrhenius temperature dependence with only a weak sensitivity to Cr composition over 1 to 9 at.\% Cr, consistent with a thermally activated migration process in this concentration range.

We further showed that the preferred chemical identity of the migrating dumbbell evolves systematically with temperature and composition, with mixed (Fe--Cr type) dumbbells dominating at low temperature and a pronounced shift towards Fe--Fe dumbbells at higher temperatures, especially in the dilute limit. The associated displacement coupling, quantified through CCA, reveals strong Fe-dumbbell coupling at higher temperatures and concentrations, whereas Cr-dumbbell coupling weakens at higher temperatures in dilute alloys. These configuration preferences and correlations provide an atomistic basis for the observed temperature-composition dependence of the defect mediated Onsager coefficients. The growth of Fe--Fe type dumbbell occupancy and strong Fe-dumbbell coupling amplify $l_{FeFe}$ and $l_{FedI}$, while reduced Cr involvement at high temperature suppresses Cr correlated transport, reflected in saturation or relative weakening of $l_{CrdI}$. Finally, in the context of RIS, our defect-species coupling data support the physical picture that the interstitial contribution to Cr enrichment is strongest in the dilute/low-temperature regime, and diminishes as increasing temperature promotes Fe--Fe dumbbell transport and weakens Cr-defect coupling, causing defect-correlated transport to become increasingly Fe dominated.

The path-resolved framework introduced here therefore provides a route for evaluating interstitial-mediated transport directly from atomistic trajectories and may be applicable to more chemically complex concentrated alloys, including high-entropy alloys, given the availability of suitable interatomic potentials. By directly sampling the evolving local chemical environments encountered by migrating defects, the approach avoids the need for explicit parameterization of configuration-dependent transport pathways and may facilitate transport calculations in systems with increasingly complex solute-defect chemistry.

\section*{Acknowledgments}
This work was funded by the Laboratory Directed Research and Development (LDRD) Program at Idaho National Laboratory under grant number 23A1070-092FP. This research made use of Idaho National Laboratory's high performance computing systems located at the Collaborative Computing Center and supported by the Office of Nuclear Energy of the U.S. Department of Energy and the Nuclear Science User Facilities under Contract No. DE-AC07-05ID14517. National Institute of Standards and Technology (NIST) work was funded solely by the United States Government.

\bibliographystyle{elsarticle-num}
\bibliography{3_references}

@Article{LAMMPS,
  author = "A. P. Thompson and H. M. Aktulga and R. Berger and 
     D. S. Bolintineanu and W. M. Brown and P. S. Crozier and
     P. J. in 't Veld and A. Kohlmeyer and S. G. Moore and T. D. Nguyen and
     R. Shan and M. J. Stevens and J. Tranchida and C. Trott and S. J. Plimpton",
  title = "{LAMMPS} - a flexible simulation tool for
     particle-based materials modeling at the 
     atomic, meso, and continuum scales",
  journal = "Comp. Phys. Comm.",
  volume =  "271",
  pages =   "108171",
  year =    "2022",
  doi = "10.1016/j.cpc.2021.108171"
}

@article{bonny2011iron,
  title={Iron chromium potential to model high-chromium ferritic alloys},
  author={Bonny, Giovanni and Pasianot, Roberto C and Terentyev, Dmitry and Malerba, Lorenzo},
  journal={Philosophical Magazine},
  volume={91},
  number={12},
  pages={1724--1746},
  year={2011},
  publisher={Taylor \& Francis}
}

@book{allnatt2003atomic,
  title={Atomic transport in solids},
  author={Allnatt, Alan R and Lidiard, Alan B},
  year={2003},
  publisher={Cambridge University Press}
}

@article{piochaud2016atomic,
  title={Atomic-based phase-field method for the modeling of radiation induced segregation in {Fe}-{Cr}},
  author={Piochaud, JB and Nastar, M and Soisson, F and Thuinet, L and Legris, A},
  journal={Computational Materials Science},
  volume={122},
  pages={249--262},
  year={2016},
  publisher={Elsevier}
}

@article{polak1969note,
  title={Note sur la convergence de m{\'e}thodes de directions conjugu{\'e}es},
  author={Polak, Elijah and Ribiere, Gerard},
  journal={Revue fran{\c{c}}aise d'informatique et de recherche op{\'e}rationnelle. S{\'e}rie rouge},
  volume={3},
  number={16},
  pages={35--43},
  year={1969},
  publisher={EDP Sciences}
}

@article{senninger2016modeling,
  title={Modeling radiation induced segregation in iron--chromium alloys},
  author={Senninger, Oriane and Soisson, Fr{\'e}d{\'e}ric and Mart{\'\i}nez, Enrique and Nastar, Maylise and Fu, Chu-Chun and Br{\'e}chet, Yves},
  journal={Acta Materialia},
  volume={103},
  pages={1--11},
  year={2016},
  publisher={Elsevier}
}

@article{terentyev2008migration,
  title={On the migration and trapping of single self-interstitial atoms in dilute and concentrated {Fe}--{Cr} alloys: Atomistic study and comparison with resistivity recovery experiments},
  author={Terentyev, D and Olsson, P{\"a}r and Klaver, TPC and Malerba, L},
  journal={Computational Materials Science},
  volume={43},
  number={4},
  pages={1183--1192},
  year={2008},
  publisher={Elsevier}
}

@article{pechenkin2013radiation,
  title={On the radiation-induced segregation: Contribution of interstitial mechanism in {Fe}--{Cr} alloys},
  author={Pechenkin, VA and Molodtsov, VL and Ryabov, VA and Terentyev, D},
  journal={Journal of nuclear materials},
  volume={433},
  number={1-3},
  pages={372--377},
  year={2013},
  publisher={Elsevier}
}

@article{einstein1905motion,
  title={On the motion of small particles suspended in liquids at rest required by the molecular-kinetic theory of heat},
  author={Einstein, Albert and others},
  journal={Annalen der physik},
  volume={17},
  number={549-560},
  pages={208},
  year={1905}
}

@article{helfand1960transport,
  title={Transport coefficients from dissipation in a canonical ensemble},
  author={Helfand, Eugene},
  journal={Physical Review},
  volume={119},
  number={1},
  pages={1},
  year={1960},
  publisher={APS}
}

@article{kuvcera1974self,
  title={Self-diffusion of iron in $\alpha$-phase of iron and {Fe}-{Cr} alloys},
  author={Ku{\v{c}}era, J and Million, B and R{\r{u}}{\v{z}}i{\v{c}}kov{\'a}, J and Foldyna, V and Jakobova, A},
  journal={Acta Metallurgica},
  volume={22},
  number={2},
  pages={135--140},
  year={1974},
  publisher={Elsevier}
}

@article{KE201883,
title = {Cluster dynamics modeling of {Mn}-{Ni}-{Si} precipitates in ferritic-martensitic steel under irradiation},
journal = {Journal of Nuclear Materials},
volume = {498},
pages = {83-88},
year = {2018},
issn = {0022-3115},
author = {Jia-Hong Ke and Huibin Ke and G. Robert Odette and Dane Morgan}
}

@article{KE2019586,
title = {Flux effects in precipitation under irradiation – Simulation of {Fe}-{Cr} alloys},
journal = {Acta Materialia},
volume = {164},
pages = {586-601},
year = {2019},
issn = {1359-6454},
author = {Jia-Hong Ke and Elaina R. Reese and Emmanuelle A. Marquis and G. Robert Odette and Dane Morgan}
}

@article{kadambi2025phase,
  title={Phase-field modeling of radiation-induced composition redistribution: An application to additively manufactured austenitic {Fe}-{Cr}-{Ni}},
  author={Kadambi, Sourabh Bhagwan and Schwen, Daniel and Ke, Jia-Hong and He, Lingfeng and Jokisaari, Andrea M},
  journal={Computational Materials Science},
  volume={255},
  pages={113895},
  year={2025},
  publisher={Elsevier}
}

@article{field2013dependence,
  title={Dependence on grain boundary structure of radiation induced segregation in a {9 wt.\% Cr} model ferritic/martensitic steel},
  author={Field, Kevin G and Barnard, Leland M and Parish, Chad M and Busby, Jeremy T and Morgan, Dane and Allen, Todd R},
  journal={Journal of Nuclear Materials},
  volume={435},
  number={1-3},
  pages={172--180},
  year={2013},
  publisher={Elsevier}
}

@article{ke2023microstructure,
  title={Microstructure modeling of nuclear structural materials: Recent progress and future directions},
  author={Ke, Jia-Hong},
  journal={Computational Materials Science},
  volume={230},
  pages={112503},
  year={2023},
  publisher={Elsevier}
}

@article{wharry2013systematic,
  title={A systematic study of radiation-induced segregation in ferritic--martensitic alloys},
  author={Wharry, Janelle P and Was, Gary S},
  journal={Journal of Nuclear Materials},
  volume={442},
  number={1-3},
  pages={7--16},
  year={2013},
  publisher={Elsevier}
}

@article{barbu1975irradiation,
  title={Irradiation-induced precipitation in {Ni}-{Si} alloys},
  author={Barbu, A and Ardell, AJ},
  journal={Scripta Metallurgica},
  volume={9},
  number={11},
  year={1975},
  publisher={CEN de Saclay, Gif-sur-Yvette, France}
}

@article{meslin2013radiation,
  title={Radiation-induced precipitation in a ferritic model alloy: An experimental and theoretical study},
  author={Meslin, E and Radiguet, Bertrand and Loyer-Prost, M},
  journal={Acta Materialia},
  volume={61},
  number={16},
  pages={6246--6254},
  year={2013},
  publisher={Elsevier}
}

@article{okamoto1974segregation,
  title={Segregation of alloying elements to free surfaces during irradiation},
  author={Okamoto, PR and Wiedersich, H},
  journal={Journal of nuclear materials},
  volume={53},
  pages={336--345},
  year={1974},
  publisher={Elsevier}
}

@article{choudhury2011ab,
  title={Ab-initio based modeling of diffusion in dilute bcc {Fe}-{Ni} and {Fe}-{Cr} alloys and implications for radiation induced segregation},
  author={Choudhury, S and Barnard, L and Tucker, JD and Allen, TR and Wirth, BD and Asta, M and Morgan, D},
  journal={Journal of Nuclear Materials},
  volume={411},
  number={1-3},
  pages={1--14},
  year={2011},
  publisher={Elsevier}
}

@article{jiao2011novel,
  title={Novel features of radiation-induced segregation and radiation-induced precipitation in austenitic stainless steels},
  author={Jiao, Z and Was, GS},
  journal={Acta Materialia},
  volume={59},
  number={3},
  pages={1220--1238},
  year={2011},
  publisher={Elsevier}
}

@article{nakata1987solute,
  title={Solute segregation along non-migrated and migrated grain boundaries during electron irradiation in austenitic stainless steels},
  author={Nakata, Kiyotomo and Masaoka, Isao},
  journal={Journal of Nuclear Materials},
  volume={150},
  number={2},
  pages={186--193},
  year={1987},
  publisher={Elsevier}
}

@article{allnatt1982einstein,
  title={Einstein and linear response formulae for the phenomenological coefficients for isothermal matter transport in solids},
  author={Allnatt, AR},
  journal={Journal of Physics C: Solid State Physics},
  volume={15},
  number={27},
  pages={5605--5613},
  year={1982}
}

@article{barnard2014ab,
  title={Ab initio molecular dynamics simulation of interstitial diffusion in {Ni}-{Cr} alloys and implications for radiation induced segregation},
  author={Barnard, L and Morgan, D},
  journal={Journal of Nuclear Materials},
  volume={449},
  number={1-3},
  pages={225--233},
  year={2014},
  publisher={Elsevier}
}

@book{flynn1972point,
  author    = {Flynn, C. P.},
  title     = {Point Defects and Diffusion},
  publisher = {Oxford University Press},
  address   = {Oxford},
  year      = {1972}
}

@article{dienes1965point,
  title={Point Defects in Metals},
  author={Dienes, GJ},
  journal={Energetics in Metallurgical Phenomena},
  volume={2},
  pages={139},
  year={1965},
  publisher={Gordon and Breach}
}

@article{olsson2007ab,
  title={Ab initio study of {Cr} interactions with point defects in bcc {Fe}},
  author={Olsson, P{\"a}r and Domain, Christophe and Wallenius, Janne},
  journal={Physical Review B—Condensed Matter and Materials Physics},
  volume={75},
  number={1},
  pages={014110},
  year={2007},
  publisher={APS}
}

@article{fu2004stability,
  title={Stability and mobility of mono-and di-interstitials in $\alpha$-{Fe}},
  author={Fu, Chu-Chun and Willaime, Fran{\c{c}}ois and Ordej{\'o}n, Pablo},
  journal={Physical review letters},
  volume={92},
  number={17},
  pages={175503},
  year={2004},
  publisher={APS}
}

@article{messina2020solute,
  title={Solute diffusion by self-interstitial defects and radiation-induced segregation in ferritic {Fe}--{X} ({X}= {Cr, Cu, Mn, Ni, P, Si}) dilute alloys},
  author={Messina, Luca and Schuler, Thomas and Nastar, Maylise and Marinica, Mihai-Cosmin and Olsson, P{\"a}r},
  journal={Acta Materialia},
  volume={191},
  pages={166--185},
  year={2020},
  publisher={Elsevier}
}

@article{barbe2007phenomenological,
  title={Phenomenological coefficients in a dilute BCC alloy for the dumbbell mechanism},
  author={Barbe, Vincent and Nastar, Maylise},
  journal={Philosophical Magazine},
  volume={87},
  number={11},
  pages={1649--1669},
  year={2007},
  publisher={Taylor \& Francis}
}

@article{vincent2006ab,
  title={Ab initio calculations of self-interstitial interaction and migration with solute atoms in bcc {Fe}},
  author={Vincent, E and Becquart, CS and Domain, C},
  journal={Journal of nuclear materials},
  volume={359},
  number={3},
  pages={227--237},
  year={2006},
  publisher={Elsevier}
}

@article{hardoon2004canonical,
  title={Canonical correlation analysis: An overview with application to learning methods},
  author={Hardoon, David R and Szedmak, Sandor and Shawe-Taylor, John},
  journal={Neural Computation},
  volume={16},
  number={12},
  pages={2639--2664},
  year={2004},
  publisher={MIT Press}
}

@incollection{hotelling1992relations,
  title={Relations between two sets of variates},
  author={Hotelling, Harold},
  booktitle={Breakthroughs in statistics: methodology and distribution},
  pages={162--190},
  year={1992},
  publisher={Springer}
}

@article{ke2022cluster,
  title={Cluster dynamics modeling of {Mn}-{Ni}-{Si} precipitates coupled with radiation-induced segregation in low-{Cu} reactor pressure vessel steels},
  author={Ke, Jia-Hong and Spencer, Benjamin W},
  journal={Journal of Nuclear Materials},
  volume={569},
  pages={153910},
  year={2022},
  publisher={Elsevier}
}

@article{nastar20121,
  title={1.18-Radiation-induced segregation},
  author={Nastar, M and Soisson, F},
  journal={Comprehensive Nuclear Materials},
  volume={1},
  pages={471--496},
  year={2012},
  publisher={Elsevier Oxford}
}

@article{wiedersich1979theory,
  title={A theory of radiation-induced segregation in concentrated alloys},
  author={Wiedersich, H and Okamoto, PR and Lam, Nghi Q},
  journal={Journal of Nuclear Materials},
  volume={83},
  number={1},
  pages={98--108},
  year={1979},
  publisher={Elsevier}
}

@article{mahmood2024universal,
  title={Universal trends in computed grain boundary energies of FCC metals},
  author={Mahmood, Yasir and Daw, Murray S and Chandross, Michael and Abdeljawad, Fadi},
  journal={Scripta Materialia},
  volume={242},
  pages={115900},
  year={2024},
  publisher={Elsevier}
}

@article{daw1984embedded,
  title={Embedded-atom method: Derivation and application to impurities, surfaces, and other defects in metals},
  author={Daw, Murray S and Baskes, Michael I},
  journal={Physical Review B},
  volume={29},
  number={12},
  pages={6443},
  year={1984},
  publisher={APS}
}

@article{Oplinger2026,
    author = {Oplinger, Russell and Bachhav, Mukesh and Ahmed, Karim and Kadambi, Sourabh Bhagwan},
    title = {Radiation-induced segregation in dilute Fe–Cr: A rate-theory framework for the Cr enrichment–depletion transition at the grain boundary},
    journal = {Journal of Applied Physics},
    volume = {139},
    number = {15},
    pages = {155903},
    year = {2026},
    month = {04},
    issn = {0021-8979},
    doi = {10.1063/5.0323296},
}

\end{document}


\begin{frontmatter}

\title{{\LARGE Supplementary Information}\\[0.5cm]
Path-Resolved Molecular Dynamics Framework for Dumbbell-Interstitial Transport in Fe--Cr Alloys}

\author[nist,umd]{Yasir Mahmood\corref{cor}}
\ead{yasir.mahmood@nist.gov}

\author[inl]{Mathew M. Swisher}

\author[inl]{Sourabh Bhagwan Kadambi}

\author[inl]{Mukesh Bachhav}

\author[inl]{Jia-Hong Ke\corref{cor}}
\ead{jiahong.ke@inl.gov}

\cortext[cor]{Corresponding authors.}

\affiliation[nist]{
organization={Materials Measurement Laboratory, National Institute of Standards and Technology},
city={Gaithersburg},
state={MD},
postcode={20899},
country={USA}
}

\affiliation[umd]{
organization={Department of Chemistry \& Biochemistry, University of Maryland},
city={College Park},
state={MD},
postcode={20742},
country={USA}
}

\affiliation[inl]{
organization={Idaho National Laboratory},
city={Idaho Falls},
state={ID},
postcode={83415},
country={USA}
}

\end{frontmatter}


\setcounter{page}{1}
\renewcommand{\thepage}{S\arabic{page}}

\renewcommand{\figurename}{Figure} 
\renewcommand{\thefigure}{S\arabic{figure}}

\renewcommand{\tablename}{Table}
\renewcommand{\thetable}{S\arabic{table}}

\section*{Onsager Transport Coefficients in the Laboratory Frame}
\begin{figure}[t!]
    \begin{center}
        \begin{tabular}{c}
            \includegraphics[width = 0.99\textwidth]{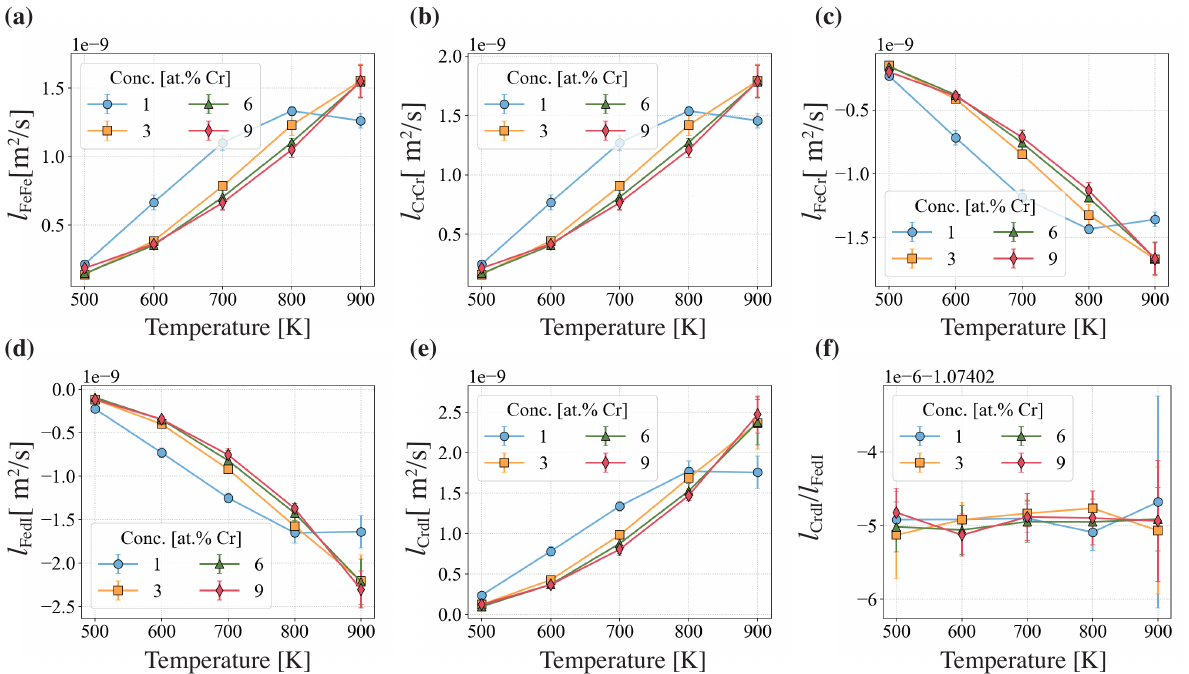}
        \end{tabular}
    \end{center}
    \caption{\label{fig:Lij_labframe} Dumbbell interstitial-mediated normalized Onsager transport coefficients in Fe--Cr using laboratory-frame, system-wide Fe and Cr collective displacements and the corresponding dumbbell displacement: (a) $l_{FeFe}$, (b) $l_{CrCr}$, (c) $l_{FeCr}$, (d) $l_{FedI}$, (e) $l_{CrdI}$, and (f) $l_{CrdI}/l_{FedI}$, shown as functions of temperature and at.\% Cr concentration. Markers denote means over ten independent runs and error bars denote standard deviations.}
\end{figure}

As a control calculation, we also evaluated the normalized Onsager coefficients, Eq.~(3) of the main text, using collective displacements of all atoms in the simulation cell, measured in the laboratory frame, without separating trajectory (T) and non-trajectory (NT) atoms over each sampling interval of 1~ns. The resulting coefficients, shown in Fig.~\ref{fig:Lij_labframe}, exhibit qualitatively different and physically inconsistent trends compared with the path-resolved calculation, including negative $l_{FeCr}$ and $l_{FedI}$ coupling terms. These artifacts indicate that the all-atom laboratory-frame displacements are dominated by the compensating system-wide lattice response rather than the localized defect-transport signal. This comparison further highlights the need to remove the background response before evaluating cross-correlation transport coefficients.
